\documentclass[11pt]{article}
\usepackage{amssymb}

\begin{document}

\title{Indirect Bounds on $Z\to \mu e,$ and\\
Lepton Flavor Violation at Future Colliders}
\author{David Del\'epine and Francesco Vissani \\
\emph{INFN, Laboratori Nazionali del Gran Sasso,}\\
\emph{Theory Group. I-67010 Assergi (AQ), Italy}}
\date{}
\maketitle

\begin{abstract}
Motivated by the interest in lepton 
number violating processes, we study
the
connection between the rate of $Z \to \mu e$ 
decay and those of the
low-energy processes $\mu \to 3 e,$ and $\mu \to e$ conversion in
nuclear
field. We show that if the vector or axial $Z$ form factors are
dominant, $Br(Z \to \mu e)$ is 
not observable, while if the $Z$ dipole form factors
are
dominant, the relatively weak indirect bound $Br(Z \to \mu e)<7\times
10^{-9}$ does not fully
preclude a signal at future colliders, as TESLA.
We finally comment on
the relation of $Z \to \mu e$ with $Z \to \tau e$ 
and $Z \to \tau \mu$
decays, and suggest a simple scaling law for these
three processes.
\end{abstract}

\setcounter{footnote}{0}
\section{Introduction}
There are convincing evidences that 
neutrinos are massive and oscillate
in
flavor. This has as natural consequence 
an increased interest in lepton
flavor violating processes, which is testified by a number of
theoretical
studies--see for instance \cite{bdv1}. Several experiments that may
considerably improve on lepton flavor violating processes are under
consideration. Quite remarkably, the ``GigaZ option'' in the TESLA
Linear
Collider project will work at the $Z$ resonance, reaching a $Z$
production
rate of $10^{9}$/year \cite{tesla}. In this way, 
TESLA could improve
by 2
or 3 orders of magnitude the LEP bounds:
\begin{eqnarray}
Br(Z\to \mu e) &<&1.7\times 10^{-6}\ \cite{opal}  \label{zmueexp} \\
Br(Z\to \tau \mu) &<&1.2\times 10^{-5}\ \cite{opal,delphi}
\label{zmutauexp}
\\
Br(Z\to \tau e) &<&9.8\times 10^{-6}\ \cite{opal,L3}  \label{ztaueexp}
\end{eqnarray}
(with $\mu e=\mu^+ e^- +\mu^- e^+,$ \textit{etc.}) or perhaps could
observe
some of these processes.

However, the question arises, whether it is possible to reconcile big
lepton-flavor violating $Z$ 
decay rates with the stringent experimental
bounds on low energy processes, such as:
\begin{eqnarray}
Br(\mu \to e\gamma ) &<&1.2\times 10^{-11}\ \cite{mega}
\label{muegamma} \\
Br(\mu \to 3e) &<&1.0\times 10^{-12}\ \cite{SINDRUM}  \label{mu3e} \\
R(\mu \to e\mbox{ in } {}^{48}_{22}{\mathrm{Ti}}) &<&6.1\times 10^{-13}\
\cite{sindrum2}  \label{conversion}
\end{eqnarray}
We denote with $R(\mu \to e$ in N)$\; = \Gamma (\mathrm{N+\mu_s \to N}
+e)/\Gamma (\mu_s $ capture in N$),$ where N is a nuclear species
($\mu_s$
=stopped muon). In near future, these results will be improved. Indeed,
at
PSI, a new experiment plans to push $\mu \to e\gamma $ down to
$10^{-14}$
\cite{Barkov}; at Brookhaven, the MECO Collaboration aims at a
sensitivity
better than $10^{-16}$ \cite{MECO} for $R(\mu \to e$ in Al).

In the present work, we obtain the 
conditions under which the constraint
on $Z\to \mu e $ 
(coming from present $\mu \to 3e$ and 
$\mu \to e$ conversion
bounds) 
are weakened. We highlight a 
special case, that could fall
within
the TESLA reach. Rather straightforwardly, we first discuss our
parametrisation of the $Z\mu e $ vertex, then describe the generic
assumptions we use, and finally study the limits on these form factors
(and
thence the indirect bounds on $Z\to \mu e $) coming from the process
$\mu
\to 3e$ and $\mu \to e$ conversion in $_{22}^{48}$Ti. We finally
discuss, in
certain specific models, the relation of $Z\to \mu e$ with the other
lepton-flavor-violating channels $Z\to \tau e$ and $Z\to \tau\mu.$

\section{$Z$ Lepton-Flavor-Violating Form-Factors}
The $Z\to \mu ^{+}e^{-}$ vertex can be parametrized in terms of 6 form
factors:
\begin{equation}
V^{\alpha }=\frac{g}{2c_{w}}\ \overline{u}_{e}(q-p)\left\{ \left( 
\gamma^{\alpha }A_{1}^{L}+\ i\sigma ^{\alpha \beta }\frac{q_{\beta
}}{M}A_{2}^{L}+
\frac{q^{\alpha }}{M}A_{3}^{L}\right) P_{L}+ (L\leftrightarrow
R)\right\}
u_{\mu }(-p)  \label{zmue}
\end{equation}
where $q$ is the $Z$ four-momentum, $g$ is the $SU(2)_{L}$ gauge
coupling
and $c_{w}=\cos \theta _{w}$ 
($\theta _{w}$=weak mixing angle). The mass
scale $M$ is introduced to make the form factors $A_{2,3}^{L,R}$
dimensionless. All form factors, in general, depend on $q^{2};$ for
on-shell
$Z$, the $A_{3}^{L,R}$ form factors do not contribute. Since $\Gamma
_{Z}^{tot} \simeq {8\times G_{F}m_{Z}^{3}}/(6\sqrt{2}\pi )$, one gets
for
the branching ratio:
\begin{equation}
Br(Z\to \mu e)\simeq \frac{1}{8}\left( \left| A_{1}^{L}\right| ^{2}
+\frac{1}{2}\left| 
\frac{m_{Z}A_{2}^{L}}{M}\right| ^{2}+ (L\leftrightarrow
R)\right)
\label{BrZmue}
\end{equation}

To proceed, we have to make certain assumptions. The first one is on the
$q^{2}$ dependence of the form factors. Indeed, processes like $\mu \to
3e$
and $\mu \to e$ conversion in $_{22}^{48}$Ti probe the form factors at
$q^{2} \sim m_{\mu }^{2}$ $(i=1,2,3).$ So our main assumption is simply
that
\begin{equation}
A_{i}^{L,R}(m_{Z}^{2})\sim A_{i}^{L,R}(m_{\mu }^{2})\simeq A_{i}^{L,R}
\label{mass}
\end{equation}
This is expected to happen if the scale of new physics $M$ responsible
for
the family lepton numbers violation is bigger than $m_{Z}$. For this
reason,
we systematically omit the $q^{2}$ dependence of the form factors. The
second assumption is about cancellations. We will suppose that the
experimental constraints are obeyed individually by the $Z$ exchange
contributions, which amounts to assume no major cancellation with other
contributions (indeed, $\mu \to 3e$ or $\mu \to e$ conversion may get
contributions, different from $Z$ bosons exchange).

\section{Indirect Bounds on $Z\to \mu e$ from $\mu \to 3e$}
The exchange of a virtual $Z$ boson leads to $\mu \to 3e$ decay. In the
limit\footnote{The 
contributions of $A_3^{L,R}$ to $\Gamma(\mu \to 3e)$ are suppressed
by a
factor $m_{e}.$ They could contribute substantially to this process (but
not
to $Z\to \mu e$) if $A_3/{M}\sim A_1 /{m_{e}},$ namely $|A_{3}|> 10^5
\times
|A_1|,$ assuming that $M > m_Z.$ We shall disregard such a possibility.}
$m_{e}\to 0,$ and with the definitions (\ref{zmue}) we get the rate:
\begin{equation}
\Gamma (\mu \to 3e) = \frac{G_{F}^{2}m_{\mu }^{5}c_{L}^{2}}{24\pi ^{3}}
\left\{ \left| A_{1}^{L}-\frac{m_{\mu }A_{2}^{R}}{2M}\right|
^{2}+\frac{1}{2}
\left| A_{1}^{R}-\frac{m_{\mu }A_{2}^{L}}{2M}\right|
^{2}+\frac{3}{40}\left|
\frac{m_{\mu }A_{2}^{L}}{M}\right| ^{2}\right\} +(L\leftrightarrow R)
\end{equation}
where $c_{L}=-1/2+s_{w}^{2}$ and $c_{R}=s_{w}^{2}$ are the usual
electron-$Z$
couplings ($s_{w}=\sin \theta _{w}$). Using the muon decay rate, and
noting
that $s_{w}^{2}\approx 1/4,$ we obtain an useful 
expression for the $\mu \to 3e$ branching 
ratio, that can be compared with eq.\ (\ref{BrZmue}):
\begin{equation}
\frac{Br(\mu \to 3e)}{Br(Z\to \mu e)}\simeq 6\times \frac{\left|
A_{1}^{L}-
\frac{m_{\mu }A_{2}^{R}}{2M}\right| ^{2}+\frac{1}{20}\left| 
\frac{m_{\mu}A_{2}^{R}}{M}\right| ^{2}+
(L\leftrightarrow R)}{\left| 
A_{1}^{L}\right|^{2}+\frac{1}{2}\left| 
\frac{m_{Z}A_{2}^{R}}{M}\right|^{2}+(L\leftrightarrow R)}  
\label{limit muee}
\end{equation}
(note the different masses, $m_\mu$ and $m_Z$ in the numerator and in
denominator respectively).

Two limiting cases are of particular interests:

\begin{enumerate}
\item  $\left| A_{1}^{L}\right| $ or $\left| A_{1}^{R}\right| \gg
\frac{m_{\mu }^{2}}{M^{2}}\left| 
A_{2}^{L,R}\right| $ $\Longrightarrow Br(\mu
\to
3e)/Br(Z\to \mu e)\simeq 6.$

One should emphasize that this case corresponds to most of unified  
theories
where usually the $Z$ dipole transitions are neglected \cite{eilam},
\cite
{GUT1}. The present experimental limit on $\mu \to 3e,$ eq.\
(\ref{mu3e}),
implies that
\begin{equation}
Br(Z\to \mu e)<1.7\times 10^{-13}  \label{3eb1}
\end{equation}
Of course, if this were the case, there would be no chance to observe
the $Z\to \mu e$ decay.

\item  $\left| A_{1}^{L,R}\right| \ll \frac{m_{\mu }^{2}}{M^{2}}\left|
A_{2}^{L,R}\right| \Longrightarrow Br(\mu \to 3e)/Br(Z\to \mu e)\simeq
\frac{18}{5}\frac{m_{\mu }^{2}}{m_{Z}^{2}}
\simeq 5\times 10^{-6}$.\newline
Using again the experimental limit 
we get\footnote{If $A_{1}^{L,R}={m_{\mu }}A_{2}^{R,L}/{2M}$, 
we can weaken the limit
given
in eq.\ (\ref{3eb2}) by a factor of $\approx 6;$ 
however, in the spirit of
our approach (no fine-tuning), this possibility is not stressed.}
\begin{equation}
Br(Z\to \mu e)<2\times 10^{-7}  \label{3eb2}
\end{equation}
This indirect bound is considerably weaker than the previous one, but
still
one order of magnitude better than the direct experimental bound.

This is the case when the bound from $\mu \to 3e$ is not incompatible
with a
large $Br(Z\to \mu e),$ namely within reach for the next generation of
colliders like TESLA. For this reason, it will be of particular interest
to
investigate whether \emph{some} predictive theory or model can fulfil
this
condition, or if such a fine-tuning for the $Z$ form factors has other
implications.
\end{enumerate}

Similar results have been recently obtained in ref.\ \cite{nussinov} using considerations based on unitarity.

\section{Indirect Bounds on $Z\to \mu e$ 
from $\mu \to e$ in Nuclear Field}
Let us pass to consider a second interesting process induced by virtual
$Z$
exchange, namely the $\mu \to e$ conversion in the nuclear field. In
this
case, one can use the non-relativistic limit for the nuclear weak
current
\cite{weinberg}, and consider the leading vectorial part. Neglecting the
electron mass, one has $q^{0}=p_{\mu }^{0}-p_{e}^{0}\to 0;$ once again,
the $A_{3}^{L,R}$ form factors are expected to give a negligible
contribution. If
the nucleus is not too heavy ($A<100$), the leading $Z$-contribution to
the $\mu \to e$ conversion rate is well approximated by
\cite{chang,bernabeu93}:
\begin{equation}
\Gamma (\mu \to e)=\frac{G_{F}^{2}m_{\mu }^{5}}{2\pi ^{2}}\frac{\alpha
^{3}Z_{eff}^{4}}{Z}Q_{W}^{2}\;|F(-m_{\mu }^{2})|^{2}\left( \left|
A_{1}^{L}+
\frac{m_{\mu }A_{2}^{R}}{M}\right| ^{2}+(L\leftrightarrow R)\right)
\label{mue conversion}
\end{equation}
where $Z$ is its number of protons. The weak charge
$Q_{W}=Z\;({1}/{2}-2\
s_{w}^{2})-{(A-Z)}/2$ in this formula shows that the process is, in
first
approximation, coherent; the deviations from perfect coherence are
quantified by the nuclear form factor $F(-m_{\mu }^{2}),$ that can be
measured by electron scattering \cite{scattering} ($\left|
F(-m_{\mu}^{2})\right| \simeq 0.54$ for $_{22}^{48}$Ti \cite{chang}).
The
parameter $Z_{eff}$ is the effective atomic charge, obtained by
averaging
the muon wave function over the nuclear density 
($Z_{eff}\simeq $ $17.6$
for
$_{22}^{48}$Ti \cite{chang}). In order to obtain $R(\mu \to e$ in
$_{22}^{48}
$Ti), one has to divide (\ref{mue conversion}) by the rate for muon
capture.
For $_{22}^{48}$Ti, $\Gamma (\mu \ \mbox{capture})=2.590\pm 0.012$
$10^{6}\mbox{ s}^{-1} $ \cite{suzuki}. Now, we can 
compare this partial rate with eq.\ (\ref{BrZmue}):
\begin{equation}
\frac{R(\mu \to e\mbox{ in }_{22}^{48}\mbox{Ti})}{Br(Z\to \mu e)}\simeq
32\times \frac{\left| A_{1}^{L}+
\frac{m_{\mu }A_{2}^{R}}{M}\right|^{2}
+(L\leftrightarrow R)}{\left| A_{1}^{L}\right|^{2}
+\frac{1}{2}\left|
\frac{m_{Z}A_{2}^{R}}{M}\right| ^{2}+(L\leftrightarrow R)}
\label{limit conversion}
\end{equation}
It is particularly interesting to consider two extreme cases:

\begin{enumerate}
\item  $\left| A_{1}^{L}\right| $ or $\left| A_{1}^{R}\right| \gg
\frac{m_{\mu }^{2}}{M^{2}}\left| 
A_{2}^{L,R}\right| \Longrightarrow R(\mu \to
e\mbox{ in }_{22}^{48}\mbox{Ti})/Br(Z\to \mu e)\simeq 32$.\newline
Using the experimental limit given in (\ref{conversion}), one gets
\begin{equation}
Br(Z\to \mu e)<2\times 10^{-14}  \label{stronger}
\end{equation}
which is one order of magnitude more stringent that eq.\ (\ref{3eb1}).
\item  $\left| A_{1}^{L,R}\right| \ll \frac{m_{\mu }^{2}}{M^{2}}\left|
A_{2}^{L,R}\right| \Longrightarrow 
R(\mu \to e\mbox{ in }_{22}^{48}\mbox{Ti})/Br(Z\to \mu e)$ 
$\simeq 64\times \frac{m_{\mu }^{2}}{m_{Z}^{2}}\simeq
8\times 10^{-5}$.\newline
Together with the present experimental limit, this 
implies\footnote{As noted 
for $\mu\to 3e,$ it is possible to weaken this indirect bound
at
the price of a fine-tuning: The (leading) coherent contribution can be
cancelled if $A_1^{L,R}=-m_\mu A_2^{R,L}/M.$} that
\begin{equation}
Br(Z\to \mu e)<7\times 10^{-9}
\end{equation}
which should be compared with (\ref{zmueexp}) and with (\ref{3eb2}).
While
this limit reduces the hopes to observe the $Z\to \mu e$ transition in
future colliders, it does not fully exclude an observation at TESLA.
\end{enumerate}

Two remarks are in order: $(i)$ The bounds coming from $\mu \to e$
conversion are stronger than the limit coming from $\mu \to 3e$,
partially
due to the fact that a $\mu \to e$ conversion is a coherent effect.
$(ii)$
More important, the experimental bound on coherent $\mu \to e$
conversion
might be soon strengthened by a factor 4 to 7 \cite{SINDRUMIIhtml};
indeed,
during the last run of the SINDRUM II in 1999, the number of muons
stopped was increased by a factor of 4.

For reader convenience, we summarize our results in table 1.
\begin{table}[t]
\begin{center}
\begin{tabular}{|c|c|c|}
\hline
& $\left| A_{1}^{L}\right| $ or $\left| A_{1}^{R}\right| \gg
\frac{m_{\mu
}^{2}}{M^{2}}\left| A_{2}^{L,R}\right| $ & $\left| A_{1}^{L,R}\right|
\ll
\frac{m_{\mu }^{2}}{M^{2}}\left| A_{2}^{L,R}\right| $ \\ \hline
$\mu \to 3e$ & $<1.7\times 10^{-13}$ & $<2.0\times 10^{-7}$ \\ \hline
$\mu \to e\mbox{ conversion}$ & $<2\times 10^{-14}$ & $<7\times 10^{-9}$
\\
\hline
\end{tabular}
\end{center}
\caption{Indirect bounds on $Br(Z\to \mu e),$ obtained from $\mu \to 3e$
and
coherent $\mu \to e$ conversion. The underlying hypotheses are discussed
in
the text.}
\end{table}

\section{On the Connection Between $Z\to \mu e,$
$Z\to \tau e$ and $Z\to \tau \mu$}
The general approach that we used for $Br(Z\to \mu e)$ can be adopted
for $Z\to \tau e$ and $Z\to \tau \mu$. 
One has simply to recall that 
$\Gamma_{\tau }\simeq 5\times 
\left( m_{\tau }/m_{\mu }\right)^{5}\Gamma _{\mu}.$
Unfortunately, the experimental 
bound \cite{PDG} on processes like $\tau
\to
3e$ or $\tau \to 3\mu $ are much less stringent than on muon decay. In
the
case of dominance of the vector and axial form factors, one gets
\begin{eqnarray*}
Br(\tau \to 3e) &<&2.9\times 10^{-6}\Longrightarrow Br(Z\to \tau
e)<2.5\times 10^{-6} \\
Br(\tau \to 3\mu ) &<&1.9\times 10^{-6} \Longrightarrow Br(Z\to \tau
\mu)<1.6\times 10^{-6}
\end{eqnarray*}
These bounds are slightly better than 
the direct experimental bounds, eqs.\ (\ref
{zmutauexp}) and (\ref{ztaueexp}). 

However, 
on theoretical basis, one may
expect a stricter relation between the 3 channels of 
lepton flavor violating $Z$
decays. Indeed, 
atmospheric neutrino observations and CHOOZ bounds \cite{atm}
suggest that the heaviest mass eigenstate has a comparable muon and tau
neutrinos component, while 
electron neutrino is some minor component.
Thence, we would expect that 
the rate of $Z\to \mu e$ is comparable to the
one of $Z\to \tau e,$ while $Z\to \tau \mu $ is bigger by some orders of
magnitude. Under this view, the results we outlined above would be of
more
general significance. Let us consider two specific models, in which
these
considerations can be formalized:

\emph{1)} Consider the massive neutrino $\nu _{3}(x)=\sum_{\ell }U_{\ell
3}\;\nu _{\ell }(x)$ ($\ell =e,\mu ,\tau $) that induces atmospheric
neutrino oscillations. It has mass $m_{3}\sim 50$ meV, and mixings
(=composition in flavor states) 
$|U_{\mu 3}|\sim |U_{\tau 3}|\sim 1/\sqrt{2}$
and $|U_{e3}|<0.15.$ 
Suppose that $\nu _{3}$ takes 
its mass mostly from the
coupling with a single right-handed neutrino $N$ as advocated in
\cite{king}:
\[
\delta \mathcal{L}=- \left\{ \ \mu _{\ell}\times (\overline{N}\; P_L\;
\nu_{\ell}) +h.c.\ \right\} -\frac{M}{2}\, \overline{N}\; N ;
\]
all parameters can be taken real. A modulus-versor
decomposition:
$\vec{\mu}=\mu \times (U_{e3},U_{\mu 3},U_{\tau 3})$ allows us to relate
the
parameters of the lagrangian with the properties of the massive neutrino
$\nu _{3},$ and to get in particular $m_{3}=\mu ^{2}/M$ (a typical seesaw
structure). The crucial 
point for us is that the mixing between light
and
heavy neutrino states, namely $\mu_{\ell }/M,$ has been related to light
neutrino mixings. In this simple model, the form factors are given by $
A_{1}(Z\to \mu e)=U_{\mathrm{e3}}\;U_{\mathrm{\mu 3}}\times (\mu
/M)^{2}\times f(M^{2}/m_{Z}^{2})$ and similar relations, $f$ being a
universal loop function. We get then:
\begin{equation}
\frac{Br(Z\to \mu e)}{Br(Z\to \tau e)} =\frac{U_{\mu 3}^{2}}{U_{\tau
3}^{2} }
\sim 1, \ \mbox{and} \ \ \frac{Br(Z\to \mu e)}{Br(Z\to \tau \mu )} =
\frac{U_{e3}^{2}}{U_{\tau 3}^{2}}\sim 2\times U_{e3}^{2}<0.04
\label{m1}
\end{equation}
it would seem that the best 
channel for experimental investigation is $Z\to \tau \mu ,$
while $Z\to \tau e$ should be unobservable. Since the
form factors are of the vector or axial type, 
 a 5 orders of magnitude 
suppression of $U_{\mathrm{e3}}^{2}$ would be needed to overcome the
limit coming from eq.(\ref{stronger}), $Br(Z\to \mu e)<2\times
10^{-14}.$ 

To our knowledge, this is the simplest way 
to argue for a connection between the rates. However 
this model fails badly to predict anything 
measurable, since the {\em amplitude} is suppressed by
$(\mu/M)^2=5\times 10^{-14}\, (m_3/\mbox{50 meV})\times (\mbox{1 TeV}/M).$ 
Note  the generality of this (decoupling) feature:
The amplitude scales as $(m_Z/M)^2$ in all models where the 
lepton-flavor violations are induced by heavy 
(singlet, right-handed) neutrinos.

\emph{2)} We then consider a more realistic possibility.
This is a supersymmetric 
$SU(5)\otimes U(1)_{F}$ model, where $U(1)_{F}$ is the flavor group. The
Froggatt-Nielsen mechanism \cite{froggatt} (namely the $U(1)_F$
selection
rules) allows us to explain the mass hierarchies of charged fermions,
but
also the large $\nu _{\mu }-\nu _{\tau }$ mixing \cite{yanagida}. In a
standard scenario, the supersymmetry breaking terms are universal at the
grand unification scale $\Lambda_{GUT};$ however, flavour violating
effects
are induced by the radiative corrections (more details in last paper of
\cite{bdv1}). 
At the electroweak scale, the left-left block of the mass
matrix of
the scalar leptons gets the contribution:\footnote{To 
correctly interpret these equations, one should recall 
the presence of not-spelled ``coefficients 
order unity'', that however are not expected to
change the order-of-magnitude estimations. 
Similar results can be obtained in other
unified models, as $SU(3)_{c}\otimes SU(3)_{L}\otimes SU(3)_{R}\otimes
U(1)_{F}$ \cite{bdv3}--see eq.\ (38) therein.}
\[
\delta m_{lij}^{2}\sim \frac{1}{8\pi ^{2}}(3 m^{2}_0+A^2) \ln
\frac{\Lambda
_{GUT}}{{M}}\epsilon ^{2a}\left(
\begin{array}{lll}
\epsilon ^{2} & \epsilon & \epsilon \\
\epsilon & 1 & 1 \\
\epsilon & 1 & 1
\end{array}
\right),\ \ \ \mbox{where }\epsilon=\frac{m_\mu}{m_\tau}
\]
$m_0,A$ are the universal supersymmetry breaking mass and trilinear
terms
(of the order of the electroweak scale) and ${M}$ is the average mass of
the
heavy neutrinos. The Froggatt-Nielsen 
parameter\footnote{Other values of $\epsilon$
are motivated and discussed in \cite{vissani}.} 
$\epsilon $ is: $\epsilon^{2} \simeq 1/300;$ 
finally, $a=0,1$ in the
cases when the neutrino Yukawa couplings are, respectively, large or
small.
By mass insertion method, one gets: $A_{1}^{R}(Z\rightarrow \mu e)\simeq
{\delta m_{l12}^{2}}/{m^2_0}\times f(m_0,m_{1/2})$ where $f$ is a
universal loop function, which 
depends on the sleptons and gaugino masses $m_0$ and
$m_{1/2}$. So, one gets:
\begin{equation}
\frac{Br(Z\to \mu e)}{Br(Z\to \tau e)} \sim 1,\ \mbox{and} \ \
\frac{Br(Z\to
\mu e)}{Br(Z\to \tau \mu )} \sim \epsilon ^{2}
\label{m2}
\end{equation}
incidentally, $\epsilon \sim \left|U_{e3}\right|$ 
in these models. Using
the
indirect bound in (\ref{stronger}), we conclude that it is very 
unlikely to
observe $Z\to \tau e$ decay at future colliders. 
But the branching ratio of
$Z\to \tau \mu $ is just 
$\sim \epsilon^{-2}$ larger; thus,
using the value of $\epsilon$ quoted above, 
we get $Br(Z\to \tau \mu )\lesssim 6\times
10^{-12}.$
Even with a generous allowance of coefficients 
order unity, this limit
remains stringent; thus, this model suggests that lepton flavor
violating $Z$
decays are not within reach.

\section{Summary and Discussion}

\emph{1.}  We obtained the connection between $Z\to \mu e$ 
and $\mu \to 3e,$ namely eq.\ (\ref{limit muee}).
We have shown that if the axial and vector $Z$
form factors are dominant,
$Br(Z\to \mu e)$ is far away from 
the sensitivity of future colliders.
Conversely, if the $Z$ dipole 
transitions are the dominant ones, the
constraint from $\mu \to 3e$ is much 
weaker, and $Br(Z\to \mu e)$ can be
as large as $10^{-7}.$

\noindent
\emph{2.}  The indirect bound on $Br(Z\to \mu e)$ 
from  coherent $\mu \to e$ conversion
(from eq.\ (\ref{limit conversion}))
is however stronger by more than 
$1$ order of magnitude (see again table 1). 
In the most optimistic assumption 
(=dominance of dipole form factors), 
the experimental bound on coherent  $\mu \to e$ conversion yields
the stringent indirect bound
$Br(Z\to \mu e)<7\times 10^{-9},$ which may be 
improved soon by a factor of $4-7.$

\noindent
\emph{3.} These indirect bounds are valid
{\em modulo} very specific cancellations between 
various contributions (or perhaps involving 
photon-exchange, box diagrams, or exotics)--see 
eq.\ (\ref{mass}) and discussion therein, and 
footnotes 2 and 3.

\noindent
\emph{4.} Our model-independent analysis
suggests a theoretical 
challenge, namely the construction of a model (theory)
where the dipole form factors are large.
Such a hypothetical model 
would be rather interesting in connection with 
lepton flavor violating processes, 
and in particular with $Z\to \mu e.$

\noindent
\emph{5.}   We have discussed the relation among
the decay channels $Z\to \mu e,$ $Z\to \tau e$ and $Z\to \tau \mu $
in certain specific models.  
Phenomenological and theoretical arguments 
(eqs.\ (\ref{m1},\ref{m2})) point to a simple scaling law:
$$
Br(Z\to \mu e)\,:\,Br(Z\to \tau e)\,:\,Br(Z\to \tau \mu )\sim
1\,:\,1\,:\,U_{e3}^{-2}  \label{Zemu/Ztaue/Ztaumu}
$$
We expect that such a relation holds for models where the 
sources of lepton flavor violations are tightly connected 
with neutrino masses.

\noindent
\emph{6.} We conclude by  remarking on the implications 
of previous relation, conjecturing its validity.
Let us assume that vector and axial $Z$ form factors 
dominate.
Due the indirect bounds on $Z\to \mu e$ discussed 
in the present work, a positive signal 
$Z\to \tau \mu$ would be related  
to very small values of 
$U_{e3}^2<10^{-5}.$ The decay $Z\to \tau e,$
instead, would be certainly too small to be observable.

\end{document}